\documentclass{PoS}
\usepackage{epsfig}
\usepackage{amsmath}   

\newcommand{\bea}{\begin{eqnarray}}
\newcommand{\eea}{\end{eqnarray}}
\newcommand{\bqa}{\begin{eqnarray}}
\newcommand{\eqa}{\end{eqnarray}}

\newcommand{\eps}{\varepsilon}
\newcommand{\bq}{ \begin {equation} }
\newcommand{\eq}{\end{equation}}
\newcommand{\be}{\begin{eqnarray}}
\newcommand{\ea}{\end{eqnarray}}

\title{%
{\small DESY 09--012 \newline SFB/CPP--09--12 \newline HEPTOOLS 08-230} \\[1.5cm]
New results for loop integrals:
\\
AMBRE, CSectors, hexagon
}

\ShortTitle{New results for loop integrals}

\author{Janusz Gluza \\
        Department of Field Theory and Particle Physics,
    Institute of Physics, \\
    University of Silesia, Uniwersytecka 4, PL-40-007 Katowice,
    Poland
\\
       E-mail: \email{gluza@us.edu.pl}}
\author{Krzysztof Kajda \\
        Department of Field Theory and Particle Physics,
    Institute of Physics, \\
    University of Silesia, Uniwersytecka 4, PL-40-007 Katowice,
    Poland
\\
       E-mail: \email{kkajda@us.edu.pl}}
\author{%
\speaker{Tord Riemann}%
\\
Deutsches Elektronen-Synchrotron, DESY,
   Platanenallee 6, 15738 Zeuthen, Germany
\\
        E-mail: \email{Tord.Riemann@desy.de}}
\author{Valery Yundin 
\\
   Deutsches Elektronen-Synchrotron, DESY,
   Platanenallee 6, 15738 Zeuthen, Germany
\\
   E-mail: \email{Valery.Yundin@desy.de}}

\abstract{%
We report on the three Mathematica packages {\ttfamily hexagon, CSectors, AMBRE}.
They are useful for the evaluation of one- and two-loop Feynman integrals with a dependence on several kinematical scales.
These integrals are typically needed for LHC and ILC applications, but also for higher order corrections at meson factories.
   {\ttfamily hexagon} is a new package for the tensor reduction of one-loop 5-point and 6-point functions with rank $R=3$ and  $R=4$, respectively; {\ttfamily AMBRE} is a tool for derivations of Mellin-Barnes representations;  {\ttfamily CSectors} is an interface for the package {\ttfamily sector\_decomposition} and allows a convenient, direct evaluation of tensor Feynman integrals. 
}

\FullConference{XII Advanced Computing and Analysis Techniques in Physics Research\\
		 November 3-7, 2008\\
		 Erice, Italy}

\begin{document}
\allowdisplaybreaks
\section{\label{intro}Introduction}
In recent years, we observe that 
higher energy,
higher luminosity,
higher precision,
more massive particles
at LHC and ILC,
 but also at low energy meson factories\footnote{For LHC and ILC, the statement is evident.
At meson factories, we have in  mind e.g. luminosity determination with Bhabha scattering  at two-loop accuracy.
For details on prospects at high-luminosity $\Phi$- and $B$-factories, see e.g. the recent mini-review \cite{Balossini:2008ht} and references quoted therein.},
lead to completely new demands on the efficient evaluation of Feynman diagrams, including:
\begin{itemize}
\item need  of some 3-point 4-point two-loop diagrams, including double boxes,
\item need of many $n$-point one-loop diagrams, 
\end{itemize}
with massive and massless particles participating, leading to complicated many-scale problems.

A few of the approaches to answer the requests will be shortly introduced, concentrating on our own activities and on publicly available  packages.
They are devoted to the evaluation of $L$-loop $n$-point Feynman integrals of tensor rank $R$,
with loop momenta $k_l$, with $E$ external legs with momenta $p_e$, and with $N$ internal lines with masses $m_i$ and propagators $1/D_i$:
\bea
  \label{bh-19a}
I^{\{\alpha\}}
&=&
\frac{e^{L\eps \gamma_E}}{(i\pi^{d/2})^L} \int \frac{d^d k_1 \ldots d^d k_L
~~N^{\alpha_1\cdots\alpha_R}}
     {D_1^{\nu_1} \ldots D_i^{\nu_i} \ldots D_N^{\nu_N}  }  ,
\\
D_i &=& \left( q_i^2-m_i^2\right)
~=~  \left[\sum_{l=1}^{L} c_i^l k_l + \sum_{e=1}^{E} d_i^e p_e \right]^2 - m_i^2~,
\eqa
where we call $d=4-2\eps$ the {generic dimension} and $\nu_i$ the {index of the propagator}.
The numerator may contain a tensor structure, e.g.:
\bqa
N &=& 1,~~~ k_1^{\alpha_1} k_1^{\beta_1},~~~k_1^{\alpha_1}  \cdots k_R^{\alpha_R}.
\eqa
There exists no general algorithm for the calculation of arbitrary  Feynman integrals, but there exists a rising number of tools, and some of them are publicly available.
In the next sections, I will comment on recent developments around three of them.

The Feynman integrals may be evaluated in quite different ways.
One may derive for them (systems of coupled) difference or differential equations, or one may seek a (minimal) basis of (scalar) master integrals, and solve only the latter ones, etc.
Often the Feynman parameter representation is useful, which replaces the $d$-dimensional momentum integrations by an appropriate number of parameter integrals.
Feynman parameters are introduced by the representation:
\bqa
\label{eq-appx114e}
\frac{1}{D_1^{\nu_1}D_2^{\nu_2}\ldots D_N^{\nu_N}} &=& 
\frac{\Gamma(\nu_1+\ldots + \nu_N)}{\Gamma(\nu_1)\ldots\Gamma(\nu_N)}
\int_0^1 dx_1\ldots\int_0^1dx_N
\frac{x_1^{\nu_1-1}\ldots x_N^{\nu_N-1}\delta(1-x_1\ldots -x_N)}{(x_1D_1+
  \ldots +x_N D_N)^{N_\nu}} , 
\eqa
with $N_\nu = \nu_1+\ldots \nu_N$.
The denominator of $I^{\{\alpha\}}$ contains, after introduction of Feynman
parameters $x_i$, the momentum dependent function $m^2$ with index-exponent $N_\nu$:
\bea
\label{bh-6}
(m^2)^{-(\nu_1+\ldots +\nu_N)} &=& (x_1D_1+
  \ldots +x_N D_N)^{-N_\nu}
~=~ (k_iM_{ij}k_j - 2Q_jk_j + J)^{-N_\nu} .
\eqa
Here $M$ is an $(LxL)$-matrix, $Q=Q(x_i, p_e)$ an $L$-vector and $J=J(x_i, m_i^2, p_{e_j}p_{e_l})$. The
$M, Q, J$ are linear in $x_i$. The momentum integration is now simple.
Shift the momenta $k$ such that $m^2$ has no linear term in ${\bar k}$,
$k = {\bar k} + (M^{-1}) Q$, and 
$m^2 =  {\bar k} M  {\bar k} -{Q M^{-1}Q + J}$.
For a scalar Feynman integral e.g. one gets:  
\bqa
\label{fint}
I &=& (-1)^{N_{\nu}}e^{L\eps \gamma_E}
\frac{\Gamma\left(N_{\nu}-\frac{D}{2}L\right)}
{\Gamma(\nu_1)\ldots\Gamma(\nu_N)} 
\int_0^1 \prod_{j=1}^N dx_j ~ x_j^{\nu_j-1}
\delta\left(1-\sum_{i=1}^N x_i\right) 
\frac{U_L(x)^{N_\nu-D(L+1)/2}}{F_L(x)^{N_{\nu}-DL/2}}
\eqa
with 
\bqa
U_L(x) &=& (\text{det}~ M), 
\\
\label{f-fun}
F_L(x) &=& (\text{det}~ M) ~\mu^2 ~=~ -(\text{det}~ M)~J + Q~\tilde{M}~ Q . 
\eqa
For one-loop functions it is:
$U_1(x) = \text{det}~ M = 1 = \sum x_i$ and so $U_1$ `disappears'.
Further, the construct $F_1(x) = -J+Q^2$ may be made  bilinear in $x_ix_j$: $F_1(x) =  - J (\sum x_i) + Q^2 = \sum A_{ij} x_i x_j$.
For tensor Feynman integrals the expressions are a little more involved, but they have the same structure: sums of rationals in the $x_i$ combined with non-integer powers of $U(x)$ and $F(x)$ \cite{Denner:2004iz,Czakon:2004wm,Gluza:2007rt}.

\section{\label{ambre}AMBRE.m}
\begin{figure}[t]
\begin{center}
\hfill
    \raisebox{7mm}{
        \includegraphics[scale=0.5]{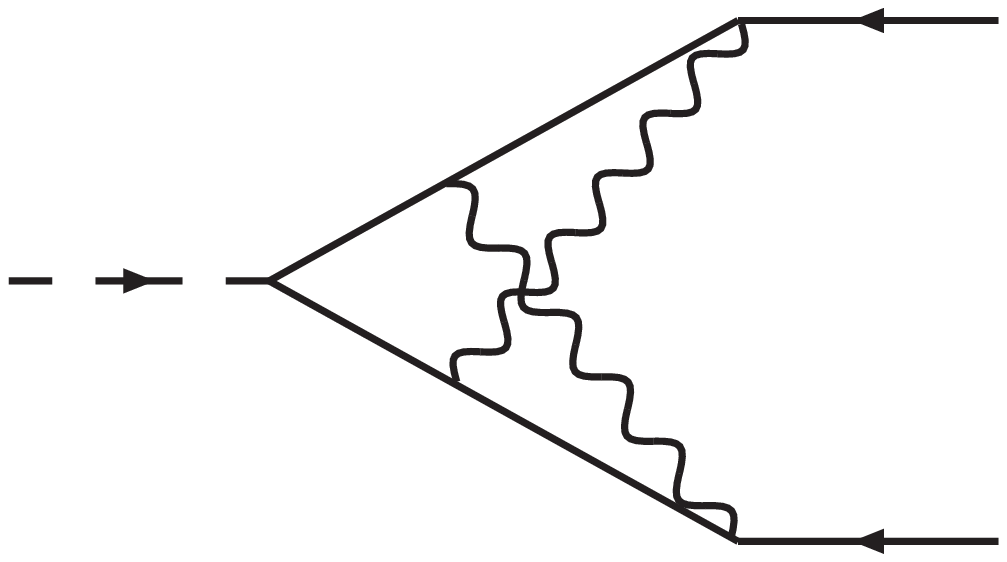}
    }
\hfill
    \includegraphics[scale=0.5]{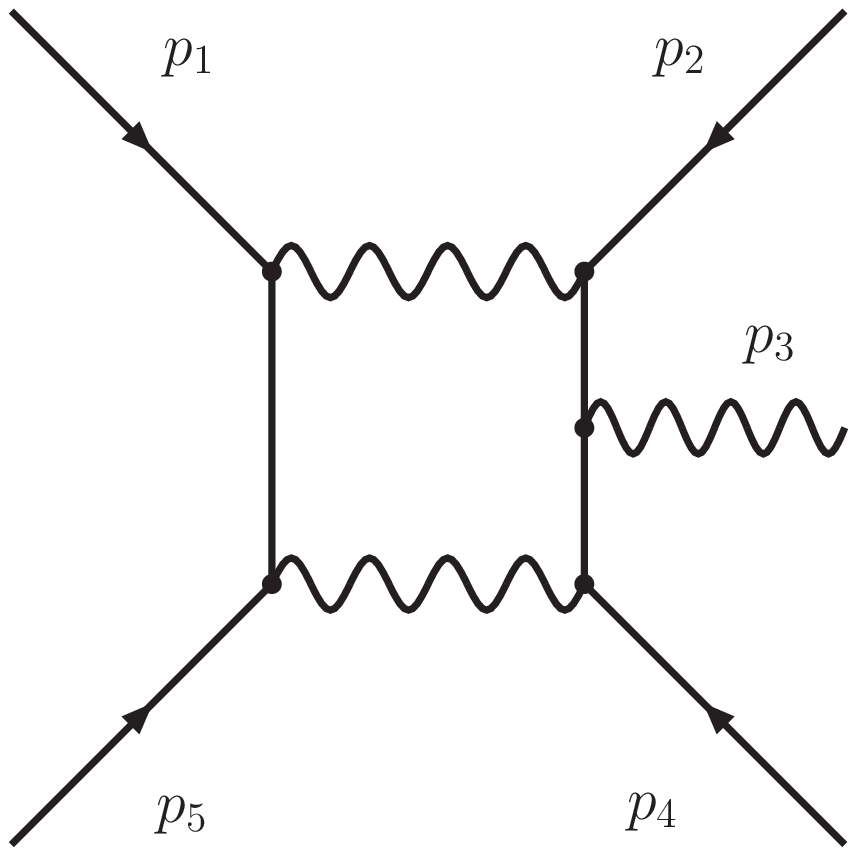}
\hfill
    \includegraphics[scale=0.5]{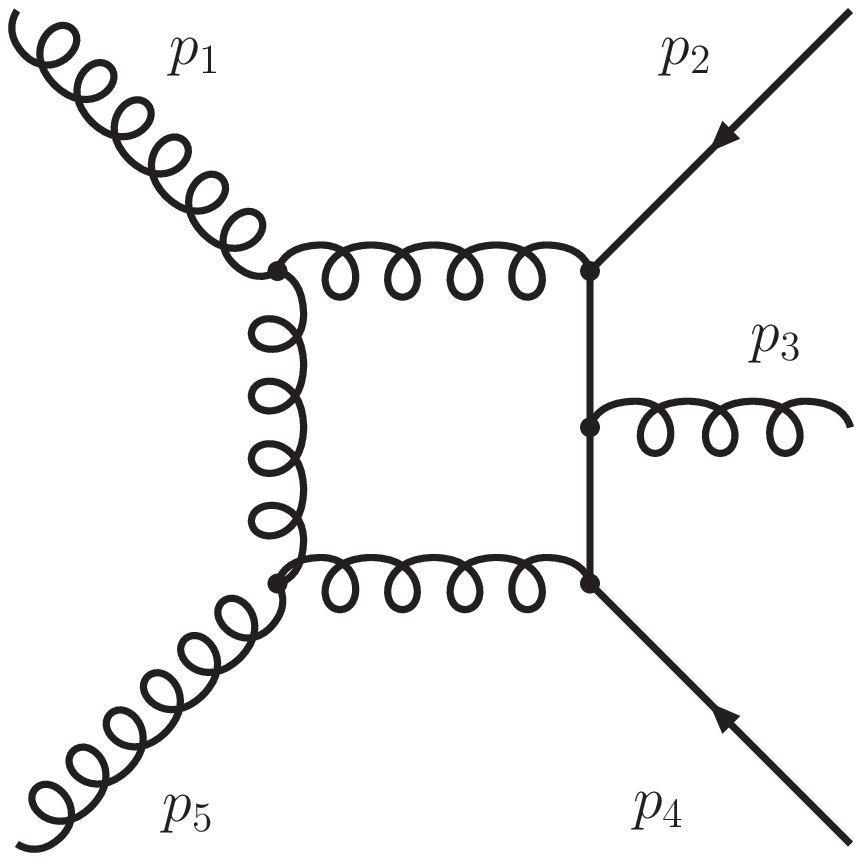}
\hfill
\caption{%
The master integrals {\ttfamily V6l4m1} and {\ttfamily F5l2m} and {\ttfamily F5l3m}. 
}
\label{fig-v6l4m}
\end{center}
\end{figure}

There is an elegant approach to the $x$-integrations.
In the $F(x)$-function (\ref{f-fun}) one may change sums of monomials in $x$ into products (as often as necessary), by Mellin-Barnes transformations, e.g.:
\bea
\frac{1}{[A(s){x_1^{a_1}} +B(t){x_1^{b_1}x_2^{b_2}} ]^a}
=
\frac{1}{2 \pi i~\Gamma(a)}
\int\limits_{-i \infty}^{i \infty}
d \sigma [A(s){x_1^{a_1}}]^\sigma
[B(t){x_1^{b_1}x_2^{b_2}}]^{a+\sigma} ~ \Gamma{(a+\sigma)}\Gamma{(-\sigma)} .
\eea
After this, one may perform the $x$-integrations:
\bqa
\label{a1}
\int_0^1 \prod_{j=1}^N dx_j ~ x_j^{\alpha_j-1}
~ \delta\left(1-\sum_{i=1}^N x_i\right) 
&=&
\frac{\Gamma(\alpha_1)~\cdots~
\Gamma(\alpha_N)}
{\Gamma\left(\alpha_1+~\cdots~
+ \alpha_N \right)} .
\eqa
Let us look  at an example, the integral {\ttfamily V6l4m1}, see Figure~\ref{fig-v6l4m}.
\footnote{The naming convention follows \cite{Czakon:2004wm}.}
In a loop-by-loop approach, after the first momentum integration one gets here $U=1$ and a first $F$-function (\ref{f-fun}), which depends yet on one internal momentum $k_1$:
 \begin{verbatim}
f1 = m^2 [X[2]+X[3]+X[4]]^2 - s X[2]X[4] - PR[k1+p1,m] X[1]X[2]
               - PR[k1+p1+p2,0] X[2]X[3] - PR[k1-p2,m] X[1]X[4]
                     - PR[k1,0] X[3]X[4] , 
\end{verbatim}
leading to a 7-dimensional MB-representation; after the second momentum integration, one has: 
\begin{verbatim}
f2 = m^2 [X[2]+X[3]]^2 - s X[2]X[3] - s X[1]X[4] - 2s X[3]X[4], 
 \end{verbatim}
leading to another 4-dimensional integral.
After several applications of Barnes' first lemma, an 8-dimensional integral has to be treated.\footnote{We made no attempt here to simplify the situation by any of the numerous tricks and reformulations etc. known to experts.}
 
The package {\ttfamily AMBRE.m} is designed for a semi-automatic derivation of Mellin-Barnes (MB) representations for Feynman diagrams; for details and examples of use see the webpage {\ttfamily http:\linebreak[3]//prac.\-us.\-edu.\-pl/$\sim$gluza/ambre/}.
The package is also available from {\ttfamily http://pro\-jects.\-hep\-forge.org/mbtools/}.
Version 1.0 is described in \cite{Gluza:2007rt}, the last released version is 1.2.
We are releasing now version 2.0, which allows to construct MB-representations for two-loop {\em tensor} integrals.
The package is yet restricted to the so-called loop-by-loop approach, which yields compact representations, but is known to potentially fail for non-planar topologies with several scales. An instructive example has been discussed in \cite{Czakon:2007wk}.   

For one-scale problems, one may safely apply {\ttfamily AMBRE.m} to non-planar diagrams.
For our example {\ttfamily V6l4m1}, one gets e.g. with the 8-dimensional MB-representation scetched above the following numerical output after running also {\ttfamily MB.m} \cite{Czakon:2005rk} (see also the webpage {\ttfamily http://pro\-jects.\linebreak[3]hep\-forge.org/mbtools/}), at $s=-11$:
\begin{eqnarray}
\label{v6l4m-mb}
\mathtt{V6l4m}~(-s)^{2\eps} = - 0.0522082~\frac{1}{\eps} - 0.17002  + 0.25606 ~\eps + 4.67 ~\eps^2 
+ {\mathcal O}(\eps^3) .
\end{eqnarray}

In simpler problems, MB-representations are a good starting point for analytical solutions, typically by summing multiple sums of residua.
Let us take as an example the  diagram {\ttfamily F5l2m} appearing in the one-loop NLO process  $gg \to q{\bar q}g$, shown in Figure~\ref{fig-v6l4m}, leading to (factorizing) two-fold infinite sums.
The steps of evaluation follow closely \cite{Gluza:2007bd,Gluza:2007uw}, and we reproduce here the result for the IR-divergent part in the infrared limit $t' = t\equiv t_m+1$, with $m=1$ and $t=(p_1+p_2)^2$, in closed form:
\begin{eqnarray}
    \mathtt{F5l2m(IR)}&=&\frac{J_{-2}}{\epsilon^2} + \frac{J_{-1}}{\epsilon} + J_0 ,
\\
J_{-2}&=&\frac{(-t_m)^{-2 \epsilon}}{s t_m^2 } 
+
    \sum_{i=1,2} 
     \frac{(v_i/s)^{ - 2 \epsilon}} {v_i}
    \frac{(-s)^{ - 2 \epsilon}}{2 s t_m} ,
\\
J_{-1} &=&  
    \frac{(-t_m)^{-2 \epsilon}}{s t_m^2 }\left[  \log(-s) + \log(v_2/s) + \log(v_4/s)\right] ,
\\
J_0&=&-\frac{\pi^2}{3}\frac{1}{s t_m} \left( \frac{1}{v_2}+\frac{1}{v_4} \right)
    -
    \frac{\pi^2}{12} 
    \frac{(-t_m)^{-2 \epsilon}}{s t_m^2 }
    +
    \sum_{i=1,2}\frac{13\pi^2}{12} 
     \frac{(v_i/s)^{ - 2 \epsilon}}{v_i}
     \frac{(-s)^{ - 2 \epsilon}}{2 s t_m} .
\end{eqnarray}
Whereever necessary, $s$ has to be replaced by $s+i \eps$.
The term $J_0$ develops infrared endpoint singularities from the phase space integrations, due to the proportionality of $v_2=p_2p_3$ and $v_4=p_4p_3$ to the gluon momentum $p_3$.

Similarly, one gets for the QED pentagon  {\ttfamily F5l3m}:
\begin{eqnarray}
    \mathtt{F5l3m(IR)}&=& \frac{J_{-1}}{\epsilon} + J_0 ,
\\
    J_{-1} &=& \sum_{i=1,2}\frac{1}{2s}
    \frac{(v_i/s)^{- 2 \epsilon}}{v_i}(-s)^{- 2 \epsilon}
    \sum_{n=0}^{\infty}
    \frac{\displaystyle(t)^n}{\displaystyle\binom{2n}{n} (2 n+1)},
\\
    J_{0}&=&\sum_{i=1,2}\frac{1}{2s}
    \frac{(v_i/s)^{- 2 \epsilon}}{v_i}(-s)^{- 2 \epsilon}
    \sum_{n=0}^{\infty}
    \frac{\displaystyle(t)^n}{\displaystyle\binom{2n}{n} (2 n+1)}
    \Bigl[- 3 S_1(n) + 2 S_1(1 + 2 n) \Bigr] .
\end{eqnarray}
The answer is less singular in $\eps$, but more complicated.
The inverse binomial sums with $S_1(n)$ and $S_1(2n+1)$ in $J_0$ are performed in \cite{Gluza:2007uw}. 
The expression  quoted there for $J_0$ contained instead of the factor $(v_i/s)^{-2\eps}$ its expansion in $\eps$, thus developing terms depending on powers of $\ln(v_i/s)$. 
This was a disadvantage notations for the subsequent regularization of the phase space integrals, so the present result is more appropriate. 
\section{\label{csect}CSectors.m}
For Euclidean kinematics, the integrand for the multi-dimensional $x$-integrations is positive semi-definite.
In numerical integrations, one has to separate the poles in $d-4$, and in doing so one has to avoid overlapping singularities.
A method for that is sector decomposition.\footnote{There are quite a few recent papers on that, e.g. \cite{Binoth:2000ps,Denner:2004iz,Bogner:2007cr}, and nice reviews are given in   \cite{Heinrich:2008si,Smirnov:2008aw}.}
The intention is to separate singular regions in different variables from each other, as is nicely demonstrated by an example borrowed from  \cite{Heinrich:2008si}: 
\begin{eqnarray}
  I &=& \int_0^1 dx \int_0^1 dy \frac{1}{x^{1+a\eps} y^{b\eps} [x+(1-x)y]}
 \nonumber 
\\ 
&=&
\int_0^1 \frac{dx}{x^{1+(a+b)\eps} }\int_0^1 \frac{dt}{t^{b\eps}[1+(1-x)t]}
+
\int_0^1 \frac{dy}{y^{1+(a+b)\eps}} \int_0^1 \frac{dt}{t^{1+a\eps}[1+(1-y)t]} .
\end{eqnarray}

At several occasions, we used the 
package {\ttfamily sector\_decomposition} \cite{Bogner:2007cr}
(built on the C++ library GINAC \cite{Bauer:2000cp}) 
 for cross checks and felt a lack of simple treatment of Feynman integrals with numerators.
For that reason, the interface {\ttfamily CSectors} was written; it will be made publicly available soon.
The syntax is similar to that of {\ttfamily AMBRE}.
The program input for the evaluation of the integral {\ttfamily V6l4m1} is simple; we again choose  $m=1, s=-11$, and the topology may be read from the arguments of propagator functions {\ttfamily PR}:

\clearpage

 \begin{verbatim}
<< CSectors.m

Options[DoSectors]
SetOptions[DoSectors, TempFileDelete -> False, SetStrategy -> C]

n1 = n2 = n3 = n4 = n5 = n6 = n7 = 1;
m = 1; s = -11;
invariants = {p1^2 -> m^2, p2^2 -> m^2, p1 p2 -> (s - 2 m^2)/2};

DoSectors[{1}, 
  {PR[k1,0,n1]       PR[k2,0,n2]       PR[k1+p1,m,n3] 
   PR[k1+k2+p1,m,n5] PR[k1+k2-p2,m,n6] PR[k2-p2,m,n7]}, 
   {k2, k1}, invariants][-4, 2]
\end{verbatim}

Here, the numerator is 1 (see the first argument $\{1\}$ of {\ttfamily DoSectors}), and the output contains the functions $U_2$ and $F_2$:
\begin{verbatim}
Using strategy C
U = x3 x4+x3 x5+x4 x5+x3 x6+x5 x6+x2 (x3+x4+x6)+x1 (x2+x4+x5+x6)

F = x1 x4^2+13 x1 x4 x5+x4^2 x5+x1 x5^2+x4 x5^2+13 x1 x4 x6
+2 x1 x5 x6+13 x4 x5 x6+x5^2 x6+x1 x6^2+x5 x6^2+x3^2 (x4+x5+x6)
+x2(x3^2+x4^2+13 x4 x6+x6^2+x3 (2 x4+13 x6))+x3 (x4^2+(x5+x6)^2
+x4 (2x5+13 x6))
\end{verbatim}
Notice the presence of a $U$-function and the complexity of the $F$-function (compared to $U=1$ and  {\ttfamily f1} and {\ttfamily f2} in the loop-by-loop MB-approach) due to the non-sequential, direct performance of both momentum integrals at once.
Both $U$ and $F$ are evidently positive semi-definite.  
The numerical result for the Feynman integral is:
\begin{eqnarray}
\label{v6l4m-cs}
\mathtt{V6l4m1}(-s)^{2\eps} =  - 0.052210 ~\frac{1}{\eps} - 0.17004 + 0.24634 ~\eps + 4.8773 ~\eps^2 
+ {\mathcal O}(\eps^3) .
\end{eqnarray}
The numbers may be compared to (\ref{v6l4m-mb}).
We obtained a third numerical result, also by sector decomposition,  with the Mathematica package  {\ttfamily FIESTA} \cite{Smirnov:2008py}:
\begin{eqnarray}
\label{v6l4m-fiesta}
\mathtt{V6l4m1}(-s)^{2\eps} =  - 0.052208 ~\frac{1}{\eps} - 0.17002 + 0.24622 ~\eps + 4.8746 ~\eps^2 
+ {\mathcal O}(\eps^3) .
\end{eqnarray}
The most accurate result can be obtained with an analytical representation based on harmonic polylogarithmic functions \cite{Remiddi:1999ew,Maitre:2005uu} obtained by solving a system of differential equations \cite{Gluza:2008-unpubl}:
\begin{eqnarray}
\label{v6l4m-hpl}
\mathtt{V6l4m1}(-s)^{2\eps} =  
- 0.0522082 ~\frac{1}{\eps} - 0.170013 + 0.246253 ~\eps + 4.87500 ~\eps^2 
+ {\mathcal O}(\eps^3) .
\end{eqnarray}
All displayed digits are accurate here. 
\section{\label{hexag}hexagon.m}
The  Mathematica package  {\ttfamily hexagon}  v. 1.0 (19 Sep 2008) was released quite recently [30 Sep 2008] at {\ttfamily http://www-zeuthen.desy.de/theory/research/CAS.html}.
It may be used  for the reduction  of one-loop tensor 5- and 6-point Feynman integrals up to ranks $R=4$ and $R=3$, respectively, to scalar 3- and 4-point Feynman integrals.
The latter have to be evaluated by some other package like {\ttfamily LoopTools}  \cite{Hahn:2006qw,Hahn:1998yk} (case of only massive internal lines) or {\ttfamily QCDloops}  \cite{Ellis:2007qk} (general case); both these packages make also use of {\ttfamily FF}   \cite{vanOldenborgh:1991yc}.
The formalism underlying this reduction and a short decsription as well as numerical examples may be found elsewhere \cite{Fleischer:2007ph,Diakonidis:2008dt,Diakonidis:2008ij,Diakonidis:0901.4455} so that we may hold this write-up short here.
We only mention that it does not use Feynman parameters, but is based on recurrence relations with dimensional shifts \cite{Fleischer:1999hq}.
In this approach, we have shown quite recently  how to cancel explicitely and completely the inverse powers of the leading Gram determinant. 
{\ttfamily hexagon} was the first publicly available tensor reduction program for 5- and 6-point Feynman integrals with arbitrary internal masses.
Now,  also {\ttfamily GOLEM95} \cite{Binoth:2008uq} became public, but in the released version it handles so far only  massless internal particles.
\section{Summary}
We described new features of the packages  {\ttfamily AMBRE} and   {\ttfamily hexagon} and of the  interface
\\
 {\ttfamily CSectors}.

\acknowledgments{%
We would like to thank T. Diakonidis, J. Fleischer and B. Tausk for fruitful collaborations in the hexagon project.
The present work is supported in part 
by the European Community's Marie-Curie Research Training Networks  MRTN-CT-2006-035505 `HEPTOOLS'
and MRTN-CT-2006-035482 `FLAVIAnet', 
and by  
Sonderforschungsbe\-reich/Trans\-regio 9--03 of Deutsche Forschungsgemeinschaft
`Compu\-ter\-gest{\"u}tzte Theo\-re\-ti\-sche Teil\-chen\-phy\-sik'. }

\providecommand{\href}[2]{#2}

\begingroup\raggedright\endgroup

\end{document}